# Revealing defect-induced spin disorder in nanocrystalline Ni

Mathias Bersweiler,[1] Evelyn Pratami Sinaga,[1] Inma Peral,[1] Nozomu Adachi,[2] Philipp Bender,[3]
Nina-Juliane Steinke,[4] Elliot Paul Gilbert,[5] Yoshikazu Todaka,[2] Andreas Michels,[1] and Yojiro Oba,[6]

[1] *Department of Physics and Materials Science, University of Luxembourg, 162A Avenue de la Faïencerie,
L-1511 Luxembourg, Grand Duchy of Luxembourg*
[2] *Department of Mechanical Engineering, Toyohashi University of Technology, 1-1 Hibarigaoka, Tempaku, Toyohashi, Aichi 441-8580, Japan*
[3] *Heinz Maier-Leibnitz Zentrum (MLZ), Technische Universität München, D-85748 Garching, Germany*
[4] *Institut Laue-Langevin, 71 avenue des Martyrs, F-38042 Grenoble, France*
[5] *Australian Centre for Neutron Scattering, Australian Nuclear Science and Technology Organization,
Locked Bag 2001, Kirrawee DC, New South Wales 2232, Australia*
[6] *Materials Sciences Research Center, Japan Atomic Energy Agency, 2-4 Shirakata, Tokai, Ibaraki 319-1195, Japan*

We combine magnetometry and magnetic small-angle neutron scattering to study the influence of the microstructure on the macroscopic magnetic properties of a nanocrystalline Ni bulk sample, which was prepared by straining via high-pressure torsion. As seen by magnetometry, the mechanical deformation leads to a significant increase of the coercivity compared to nondeformed polycrystalline Ni. The neutron data reveal a significant spin-misalignment scattering caused by the high density of crystal defects inside the sample, which were created by the severe plastic deformation during the sample preparation. The corresponding magnetic correlation length, which characterizes the spatial magnetization fluctuations in real space, indicates an average defect size of 11 nm, which is smaller than the average crystallite size of 60 nm. In the remanent state, the strain fields around the defects cause spin disorder in the surrounding ferromagnetic bulk, with a penetration depth of around 22 nm. The range and amplitude of the disorder is systematically suppressed by an increasing external magnetic field. Our findings are supported and illustrated by micromagnetic simulations, which, for the particular case of nonmagnetic defects (holes) embedded in a ferromagnetic Ni phase, further highlight the role of localized spin perturbations for the magnetic microstructure of defect-rich magnets such as high-pressure torsion materials.

## 1. Introduction

Ultrafine-grained and nanocrystalline magnetic materials have attracted considerable interest over the last decades owing to their large potential for technological applications [1–4]. Among the most well-known and efficient techniques for synthesizing such materials are inert-gas condensation and high-pressure torsion (HPT), the latter being a severe plastic deformation method. For a brief overview of the main techniques for the preparation of bulk ultrafine-grained and nanocrystalline materials, we refer the reader to the article by Koch [5]; a detailed review on the HPT technique can be found in Ref. [6].

Since the magnetic properties will ultimately determine the performance of ultrafine-grained materials, a precise knowledge of the relationship between the microstructure and the magnetic properties, *i.e.*, the correlation between *e.g.* the saturation magnetization, coercive field, and magnetic anisotropy and the average grain size or crystallographic texture is crucial. Previous studies have reported that the magnetic properties of strained nanocrystalline materials produced by HPT strongly differ from polycrystalline samples with larger grain sizes; in particular, a reduction of the saturation magnetization (by 5 %) and a significant increase of the coercive field (∼ 50 times larger) were observed in HPT Ni using magnetometry [7]. These features were qualitatively explained, respectively, by the decrease of the exchange energy in the vicinity of defects [8] and by the increase of the dislocation density within the grain boundaries [7]. The investigation of the magnetic domain structure of HPT materials with the aim to clarify the magnetization reversal mechanism has been mainly performed using Lorentz electron microscopy [9,10]. Although these studies reported that the domain structure (*i.e.,* shape and size) is not strongly affected by the grain size, the influence of a high density of lattice defects (*e.g.* vacancies, dislocations, grain boundaries, pores) induced by HPT on the spin structure still needs to be further clarified. As previously demonstrated, HPT can be used to modify the structure and thus to control the *macroscopic* magnetic properties of magnetic materials [11,12]. Therefore, in the context of defect engineering of advanced materials using severe plastic deformation [13], a better understanding of the influence of the defects on the magnetic properties at different length scales is necessary.

In this paper we employ unpolarized magnetic small-angle neutron scattering (SANS) to investigate the magnetic microstructure of HPT Ni on the *mesoscopic* length scale. Magnetic SANS is a powerful technique which provides volume-averaged information about the perturbation of the magnetization vector field on a length scale of about 1 – 500 nm (see Refs. [14,15] for reviews of the magnetic SANS fundamentals and applications). This technique was recently used to demonstrate that in HPT Fe defects act as a source of an anomalous effective magnetic anisotropy field [16]. Here, we go a step further in the neutron data analysis. We determine the real-space magnetic correlation lengths from the magnetic SANS data to obtain estimates for the average defect size as well as for the spatial extent of the surrounding spin disorder within the bulk of the sample. Our experimental results are supported and illustrated by micromagnetic simulations. The specific neutron data analysis of the spin misalignment brings additional information which is important for the understanding of the role played by the defects in magnetic materials.

**2. Experimental details**

The HPT process for the preparation of the strained Ni sample used in this study is similar to the one described in Ref. [17]. Wide-angle x-ray diffraction data on HPT Ni and a nondeformed Ni sample (nd Ni) has been taken on a Bruker D8 diffractometer in Bragg-Brentano geometry using Cu K$_\alpha$ radiation. Room-temperature magnetization curves for both samples were recorded using a Cryogenic Ltd. vibrating sample magnetometer (VSM) equipped with a 14 T superconducting magnet. For the neutron experiments two disks of HPT Ni with a diameter of 10 mm and a thickness of 0.5 mm were stacked together, resulting in a total sample thickness of 1.0 mm. The neutron measurements were conducted at the instrument D33 at the Institut Laue-Langevin, Grenoble [18]. The measurements were done using an unpolarized neutron beam with a mean wavelength of $\lambda = 4.6$ Å and a wavelength broadening of $\Delta\lambda/\lambda = 10\%$ (full width at half maximum). The measurements were performed at room temperature and within a $q$-range of about $0.04$ nm$^{-1} \leq q \leq 0.45$ nm$^{-1}$. A magnetic field $\boldsymbol{H}_0$ was applied perpendicular to the incident neutron beam ($\boldsymbol{H}_0 \perp \boldsymbol{k}_0$). Neutron data were recorded by decreasing the field from the maximum field available of 6.7 T to 0.1 T. The neutron-data reduction (correction for the empty sample holder and background scattering, sample transmission, and detector efficiency) was conducted using the GRASP software package [19]. Further neutron experiments under similar conditions were performed at the QUOKKA instrument [20] at the Australian Nuclear Science and Technology Organization (ANSTO). We determined the (over $2\pi$) azimuthally-averaged purely magnetic SANS cross section d$\Sigma_{\text{mag}}(q, H_0)$/d$\Omega$ by subtracting the total (nuclear + magnetic) SANS cross section d$\Sigma$/d$\Omega$ measured at the highest field of 6.7 T (approaching saturation) from the d$\Sigma$/d$\Omega$ measured at lower fields. This specific neutron data analysis was previously used to study the magnetization profile of magnetic nanoparticles [21].

**3. Magnetic SANS Analysis**

*3.1. Unpolarized SANS cross section*

When the applied magnetic field is perpendicular to the incident neutron beam ($\boldsymbol{H}_0 \perp \boldsymbol{k}_0$), the elastic total (nuclear + magnetic) unpolarized SANS cross section d$\Sigma$/d$\Omega$ at momentum-transfer vector $\boldsymbol{q}$ can be written as [14, 15]:

$$\frac{d\Sigma}{d\Omega}(\boldsymbol{q}) = \frac{8\pi^3}{V} b_{\text{H}}^2 \left( b_{\text{H}}^{-2} |\widetilde{N}|^2 + |\widetilde{M}_x|^2 + |\widetilde{M}_y|^2 \cos^2(\theta) \right.$$
$$+ |\widetilde{M}_z|^2 \sin^2(\theta)$$
$$\left. - (\widetilde{M}_y \widetilde{M}_z^* + \widetilde{M}_y^* \widetilde{M}_z) \sin(\theta) \cos(\theta) \right) \quad (1)$$

where $V$ is the scattering volume, $b_H = 2.91 \times 10^8$ A$^{-1}$m$^{-1}$ relates the atomic magnetic moment to the atomic magnetic scattering length, $\widetilde{N}(\boldsymbol{q})$ and $\widetilde{\boldsymbol{M}}(\boldsymbol{q}) = [\widetilde{M}_x(\boldsymbol{q}), \widetilde{M}_y(\boldsymbol{q}), \widetilde{M}_z(\boldsymbol{q})]$ represent the Fourier transforms of the nuclear scattering length density $N(\boldsymbol{r})$ and of the magnetization vector field $\boldsymbol{M}(\boldsymbol{r})$, respectively, $\theta$ specifies the angle between $\boldsymbol{H}_0$ and $\boldsymbol{q} \cong q\{0, \sin\theta, \cos\theta\}$ in the small-angle approximation, and the asterisks "*" denote the complex conjugated quantities. For small-angle scattering, the component of the scattering vector along the incident neutron beam, here $q_x$, is smaller than the other two components $q_y$ and $q_z$, so that only correlations in the plane perpendicular to the incoming neutron beam are probed.

In our neutron-data analysis below, we subtract the SANS signal at the largest available field of 6.7 T [approach-to-saturation regime; compare Fig. 1(c)] from the measured data at lower fields. This subtraction procedure eliminates the nuclear SANS contribution $\propto |\widetilde{N}|^2$, which is field independent, and it yields the following purely magnetic SANS cross section d$\Sigma_{\text{mag}}$/d$\Omega$:

$$\frac{d\Sigma_{\text{mag}}}{d\Omega}(\boldsymbol{q}) = \frac{8\pi^3}{V} b_{\text{H}}^2 \left( \Delta|\widetilde{M}_x|^2 + \Delta|\widetilde{M}_y|^2 \cos^2(\theta) \right.$$
$$+ \Delta|\widetilde{M}_z|^2 \sin^2(\theta)$$
$$\left. - \Delta(\widetilde{M}_y \widetilde{M}_z^* + \widetilde{M}_y^* \widetilde{M}_z) \sin(\theta) \cos(\theta) \right) , (2)$$

where the "$\Delta$" stand for the differences of the Fourier components at the two fields considered. We emphasize that d$\Sigma_{\text{mag}}$/d$\Omega$ is strongly dominated by the two transversal magnetization Fourier components $\widetilde{M}_{x,y}$.

*3.2. Magnetic correlation function*

The normalized magnetic correlation function $C(r, H_0)$ was numerically computed by a direct Fourier transformation of the experimental data for d$\Sigma_{\text{mag}}(q, H_0)$/d$\Omega$ according to [22]:

$$C(r, H_0) = \frac{\int_0^\infty \frac{d\Sigma_{\text{mag}}(q, H_0)}{d\Omega} j_0(qr) q^2 dq}{\int_0^\infty \frac{d\Sigma_{\text{mag}}(q, H_0)}{d\Omega} q^2 dq} , \quad (3)$$

where $j_0(qr) = \sin(qr)/qr$ is the zeroth-order spherical Bessel function. For this purpose, the experimental data d$\Sigma_{\text{mag}}(q, H_0)$/d$\Omega$ beyond $q_{\text{max}}$ were extrapolated to infinity using a power law, d$\Sigma_{\text{mag}}(q, H_0)$/d$\Omega \propto 1/q^n$ with $4 \lesssim n \lesssim 7$ [compare inset in Fig. 2(b)], and the extrapolation from $q_{\text{min}}$ to $q = 0$ was done according to d$\Sigma_{\text{mag}}(q, H_0)$/d$\Omega \propto a + bq^2$.

*3.3. Magnetic correlation length*

The magnetic correlation length $l_c$ characterizes the distance over which perturbations in the spin structure around a lattice defect are transmitted by the exchange interaction into the surrounding crystal lattice [23,24]. Several procedures for obtaining $l_c$ are discussed in the literature, e.g., $l_c(H_0)$ can be defined as the value of $r$ for which $C(r, H_0) = C(0)e^{-1}$, or $l_c$ can be found from the logarithmic derivative of $C(r, H_0)$ in the limit $r \to 0$ (Ref. [23]). Here, we determined the magnetic

correlation length $l_c(H_0)$ from the $C(r,H_0)$ data at a particular field according to:

$$l_c(H_0) = \frac{\int_0^\infty r C(r,H_0) dr}{\int_0^\infty C(r,H_0) dr} \quad . \quad (4)$$

The field dependence of the $l_c$ data was then fitted using the following expression:

$$l_c(H_0) = \mathcal{L} + \sqrt{\frac{2 A_{ex}}{\mu_0 M_s (H_0 + H^*)}} \quad , \quad (5)$$

where the field-independent parameter $\mathcal{L}$ is of the order of the defect size, and $A_{ex}$ and $M_s$ are the exchange-stiffness constant and the saturation magnetization, respectively. The field $H^*$ models the contribution of the magnetostatic and magnetic anisotropy field to the internal magnetic field. In the approach-to-saturation regime and for single-phase materials, where nonzero divergences of the magnetization from within the bulk are expected to be small [25], the main contributions to $H^*$ are due to the magnetic anisotropy. The phenomenological model, Eq. (5), is based on micromagnetic theory [26,27] and expresses the relationship between the nuclear and magnetic microstructure of a material. Equation (5) has already been successfully used to describe the spin misalignments in several nanocrystalline bulk ferromagnetic materials [28,29]. To reduce the number of free parameters in the data analysis (see below), we fixed $M_s$ to the value estimated from the magnetization curve (see Table 1), and $A_{ex}$ to 8.5 pJ/m. This value is obtained by performing a global fit analysis of the field-dependent $d\Sigma/d\Omega$ data shown in Fig. 2(a) based on the micromagnetic SANS theory developed in Refs. [24,30]. The resulting volume-averaged value of $A_{ex} = 8.5 \pm 0.2$ pJ/m compares favorably with data reported for nanocrystalline Ni [27].

## 4. Results and discussion

Figure 1 displays the results of the structural and magnetic characterization. In Fig. 1(a) we show wide-angle x-ray diffraction data of HPT Ni and of a nondeformed Ni reference sample. The XRD patterns exhibit no impurity peaks, which confirm the high-quality synthesis of the samples. The XRD data refinement using the Le Bail fit method (LBF) implemented in the Fullprof software [31] yields an estimate for the average crystallite size of 60 ± 5 nm and a value of ≈ 0.12 % for the root mean square strain of HPT Ni. The average crystallite size and strain of the nondeformed sample could not be evaluated, since the widths of the Bragg peaks cannot be distinguished from the instrumental resolution function. Figure 1(b) and (c) present the room-temperature magnetization curves of HPT and nondeformed Ni on a linear and semi-logarithmic scale, respectively. From the magnetization curves, we estimated the saturation magnetization $M_s$, the remanent magnetization $M_r$ and the coercive field $H_c$ (see Table 1). By comparing the hysteresis curves, the two following features are observed: (*i*) $M_s$ of the HPT sample is slightly reduced (2.3%) and (*ii*) its $H_c$ increases significantly. These features are consistent with previous magnetic studies on HPT Ni [7,8,12]. The increase in $H_c$ can be attributed to the pinning of domain-wall motion by the high-density of defects and the enhanced magnetoelastic coupling energy due to HPT straining [7,12]. The decrease in $M_s$ is generally explained by the formation of a grain-boundary phase with a lower crystal symmetry and smaller $M_s$ [8]. Defining the approach-to-saturation regime by $M/M_s > 90$ %, we see from Fig. 1(c) that this regime is attained for applied fields larger than about 0.1 T.

Figure 2(a) displays the (over $2\pi$) azimuthally-averaged total (nuclear + magnetic) SANS cross section $d\Sigma(q,H_0)/d\Omega$ measured at different applied magnetic fields. As can be seen, at the smaller momentum transfers $q$ the cross section $d\Sigma(q,H_0)/d\Omega$ increases by more than two orders of magnitude when $H_0$ is decreased from 6.7 T to 0.1 T. Since the nuclear scattering is field independent, the strong field dependence of $d\Sigma(q,H_0)/d\Omega$ observed in Fig. 2(a) originates from spin-misalignment scattering caused by mesoscale spin disorder (*i.e.* from the failure of the spins to be completely aligned along $\boldsymbol{H}_0$). Figure 2(b) shows the corresponding purely magnetic SANS cross sections $d\Sigma_{mag}(q,H_0)/d\Omega$, which were obtained by subtracting the total scattering at 6.7 T (approaching saturation) from the data at lower fields. The magnitude of $d\Sigma_{mag}(q,H_0)/d\Omega$ is of the same order as $d\Sigma(q,H_0)/d\Omega$. The asymptotic power-law exponent $n$ in $d\Sigma_{mag}(q,H_0)/d\Omega \propto 1/q^n$ was found to be larger than the value of $n = 4$ [see inset in Fig. 2(b)]; $n = 4$ would correspond to scattering from particles with sharp interfaces or from exponentially correlated fluctuations. The finding of a field-dependent $n > 4$ supports the notion of dominant spin-misalignment scattering, for which exponents $n = 4–8$ are theoretically predicted and experimentally found [14,28].

Figure 3 shows the normalized magnetic correlation function $C(r,H_0)$, which was numerically computed according to Eq. (3). Increasing the field from 0.1 to 4 T results in a decrease of $C(r,H_0)$ at a given $r$. This observation reflects the decrease of the spin-misalignment fluctuations and the suppression of the amplitude of the static disorder with increasing field. Furthermore, the correlations do not decay exponentially (see the log-linear plot in the inset of Fig. 3), in agreement with the absence of a corresponding $n = 4$ power-law exponent observed in Fig. 2(b). Moreover, for the lowest applied fields, the absence of a finite slope of $C(r,H_0)$ in the limit $r \to 0$ is consistent with the absence of sharp interfaces in the magnetic microstructure and with the presence of a continuous magnetic scattering length density variation. By contrast, for homogeneous particles with a sharp interface (*i.e.*, with a discontinuous jump in the scattering-length density), the slope of the correlation function at the origin is finite and provides information on the fine structure of the particle (*e.g.*, on the surface-to-volume ratio). This is a direct consequence of the asymptotic $q^{-4}$ Porod behavior of the SANS cross section (see the discussion by Porod in Ref. [32]). As shown in Ref. [22], for bulk ferromagnets, which are characterized by a smoothly-varying continuous magnetization vector field, the slope of $C(r)$ vanishes as $r \to 0$, and the asymptotic power-law behavior of the magnetic SANS cross section exhibits power-law exponents larger than $n = 4$.

Figure 4 presents the field dependence of the magnetic correlation length $l_c(H_0)$ determined from the $C(r, H_0)$ using Eq. (4). As can be seen, $l_c(H_0)$ increases from about 14 nm at the highest field of 4 T to 26 nm at the lowest field of 0.1 T. Moreover, for all fields investigated, the values of $l_c(H_0)$ remain much smaller than the average crystallite size of 60 nm of the HPT Ni sample. This latter observation thus indicates the presence of spin-misalignment correlations on a scale smaller than the average crystallite size. From the nonlinear least squares fit of the $l_c(H_0)$ data to Eq. (5) (dashed line in Fig. 4), the following best-fit parameters are obtained: $\mathcal{L} = 11.3 \pm 0.1$ nm and $\mu_0 H^* = 71.2 \pm 3.0$ mT. We reemphasize that $\mathcal{L}$ can be regarded as an estimate of the average defect size and $H^*$ models the influence of the magnetostatic and magnetic anisotropy field contributions to the internal magnetic field. The estimated "defect size" $\mathcal{L} \sim 11$ nm suggests that the origin of the spin misalignment observed in HPT Ni results from a high density of crystal defects on a scale smaller than the grain size, as previously suggested in HPT Fe [16]. In the remanent state, we estimate the penetration depth $\delta = l_c(H_0 = 0) - \mathcal{L}$ of the spin disorder into the ferromagnetic Ni-phase to be ~ 22 nm (Fig. 4). If the field $H^*$ in Eq. (5) were exclusively due to an *effective* uniaxial magnetic anisotropy of strength $K_u^{\text{eff}}$, then $H^* = 2K_u^{\text{eff}}/(\mu_0 M_s)$ and the penetration depth $\delta = l_c(H_0 = 0) - \mathcal{L}$ would be related to the domain-wall width $\delta_w \propto \sqrt{A_{\text{ex}}/|K_u^{\text{eff}}|}$. Using $A_{\text{ex}} = 8.5$ pJ/m and assuming the (magnetocrystalline) single-crystal value of $K_u^{\text{eff}} = -5.7$ kJ/m$^3$ [33] (ignoring that fcc Ni has a cubic rather than an uniaxial anisotropy symmetry), we obtain $\delta_w \cong 39$ nm. On the other hand, using $A_{\text{ex}} = 8.5$ pJ/m and demanding that $\sqrt{A_{\text{ex}}/K_u^{\text{eff}}} = 22$ nm, as is experimentally found, we estimate $K_u^{\text{eff}} \cong 1.8 \times 10^4$ J/m$^3$. This corresponds to an increase by a factor of about 4 compared to the (magnetocrystalline) value reported in Ni single crystal. In line with the HPT process and with the finding of an increased coercivity, the above value of $K_u^{\text{eff}}$ indicates a significant contribution to the magnetic anisotropy due to magnetoelastic effects; this hypothesis is consistent with the strain observed in the HPT sample (see Table 1). The value of $2K_u^{\text{eff}}/M_s$ self-consistently evaluates to $\mu_0 H^* \cong 73$ mT (using $M_s = 482$ kA/m [taken from Table 1 and assuming a mass density of 8.912 g/cm$^3$]), in agreement with the experimental result.

To support and graphically illustrate our experimental findings, we simulated the real-space distribution of the magnetization vector field $\mathbf{M}(\mathbf{r})$ around a spherical nonmagnetic defect embedded in a ferromagnetic Ni-phase (magnetic hole or pore). The open-source software package MuMax3 (Ref. [34]) was used for this purpose. Figure 5 displays the simulation results for two selected applied fields, namely 0.05 T and 2 T. As can be seen, a localized perturbation of the magnetization (top panel) is observed around the defect at 0.05 T. This magnetization inhomogeneity is caused by the magnetodipolar stray field which is related to the jump of the magnetization magnitude at the pore-matrix interface ($\mu_0 \Delta M = 0.62$ T). In our experimental results the spatial extent of such a magnetization gradient is (at a given field) represented by $l_c$. An applied field of 2 T largely suppresses the stray-field-torque related spin disorder. This can be seen in the bottom panel of Fig. 5, which compares the perpendicular component $M_y$ at both fields. The scenario which is displayed in Fig. 5 illustrates how the nanoscale magnetization inhomogeneity, which is associated with a particular lattice defect, is related to a contrast for magnetic SANS. In real HPT samples, in addition to magnetostatic stray fields around voids, the magnetic anisotropy field generated by other microstructural defects such as vacancies, dislocations, and grain boundaries may contribute to the spin-misalignment scattering [35,36].

Computing the correlation function and correlation length from such numerical micromagnetic calculations and to compare them to experimental results would be desirable. However, this requires a more realistic description of the defect structures in micromagnetic codes. In the present HPT Ni sample, the dominating defects are dislocations, which are at the origin of the spin disorder seen by SANS. At the moment, such line defects cannot be adequately modeled in the micromagnetic framework, since this would require (among other things) the knowledge and implementation of the strain tensor.

To systematically correlate the macroscopic magnetic properties (*e.g.*, the coercivity) to the outcome of the neutron-data analysis (*e.g.*, defect size, range of the correlation) one should perform a series of measurements for different degrees of deformation (HPT straining). Such experiments are underway and could further establish the SANS technique as an important pillar for the characterization of ultrafine-grained materials in the bulk and on the relevant mesoscopic length scale.

A final comment relates to the role of unpolarized versus polarized neutrons for the study of magnetic materials. The present results clearly demonstrate that for nanocrystalline defect-rich materials, which exhibit a strongly field-dependent SANS signal, it is not necessary to resort to polarized neutrons, if the aim is to study the (polarization-independent) spin-misalignment scattering. Polarized neutrons only provide additional information via the chiral scattering term in the cross section. A recent paper [37] suggests that the local symmetry-breaking at defect sites gives rise to an asymmetric polarization-dependent contribution to the SANS cross section. In this respect, it would of interest to carry out half-polarized SANS experiments or even a one-dimensional polarization analysis on HPT samples.

## 5. Conclusion

We employed magnetometry and unpolarized magnetic SANS to investigate the influence of crystal defects on the magnetic microstructure and the macroscopic magnetic properties of nanocrystalline Ni prepared by HPT. The magnetometry data confirms a significant change of the macroscopic magnetic properties of the deformed sample (*i.e.,* a slight decrease of $M_s$ and a strong increase of $H_c$) which can be attributed to the induced structural defects. The analysis of the field-dependent magnetic SANS data suggests the presence of strong spin misalignment on the mesoscopic length scale. In fact, the computation of the magnetic correlation function and the correlation length confirmed the presence of spin disorder on a scale smaller than the average crystallite size of 60 nm. The phenomenological model, Eq. (5), provides an excellent description of the field dependence of the spin-misalignment correlation length. We estimated the defect size to be around 11 nm and the penetration depth of the spin misalignment into the pure Ni phase in the remanent state to be around 22 nm. Under certain assumptions, the analysis of the penetration depth allows one to conclude on the magnitude of the effective magnetic anisotropy. For HPT Ni, we find an increased anisotropy (factor of 4), presumably due to magnetoelastic interactions, relative to the single-crystalline ground state. Our findings are supported by micromagnetic simulations which highlight that microstructural defects (such as pores) can induce significant nanoscale spin disorder, representing a contrast for magnetic SANS. The presented neutron-data analysis procedure is particularly useful for defect-rich materials and may pave the way to tune magnetic properties through defect engineering as proposed *e.g.* for magnetic nanostructured materials such as nanoparticles [38].

**Acknowledgement**
The authors acknowledge the Institut Laue-Langevin, Grenoble, France, and the Australian Nuclear Science and Technology Organization, Lucas Heights, Australia, for the provision of neutron beamtime. The authors thank Dr. Robert Cubitt for the technical support during the neutron experiments at the Institut Laue-Langevin. Andreas Michels and Evelyn Pratami Sinaga acknowledge financial support from the National Research Fund of Luxembourg (Pride MASSENA Grant No. 10935404). Yojiro Oba thanks KAKENHI (Grant No. 19K05102), the Japan Science and Technology Agency (JST) (Grant No. JPMJSK1511), and the travel expenses support of the Institute for Solid State Physics, University of Tokyo.

**Data availability**
The data that support the findings of this study are available from the corresponding author upon reasonable request.

# Table 1

| Parameters | HPT Ni | nd Ni | Units |
|---|---|---|---|
| $D$ | $60 \pm 5$ | $> 100$ | nm |
| $\langle \varepsilon^2 \rangle^{1/2}$ | 0.12 % | – | – |
| $M_\text{s}$ | $54.1 \pm 0.1$ | $55.4 \pm 0.1$ | Am$^2$/kg |
| $M_\text{r}$ | $5.5 \pm 0.1$ | $0.3 \pm 0.1$ | Am$^2$/kg |
| $\mu_0 H_\text{c}$ | $5.0 \pm 0.1$ | $\approx 0$ | mT |

Table 1: Structural and magnetic parameters of HPT Ni and nondeformed Ni (nd Ni). The average crystallite size ($D$) and the root mean square strain ($\langle \varepsilon^2 \rangle^{1/2}$) have been determined from the XRD data refinement using the Le Bail fit method implemented in the Fullprof software. The saturation magnetizations ($M_\text{s}$), remanent magnetizations ($M_\text{r}$), and the coercive fields ($H_\text{c}$) have been estimated from the magnetization curves shown in Fig. 1(b).

Figure 1

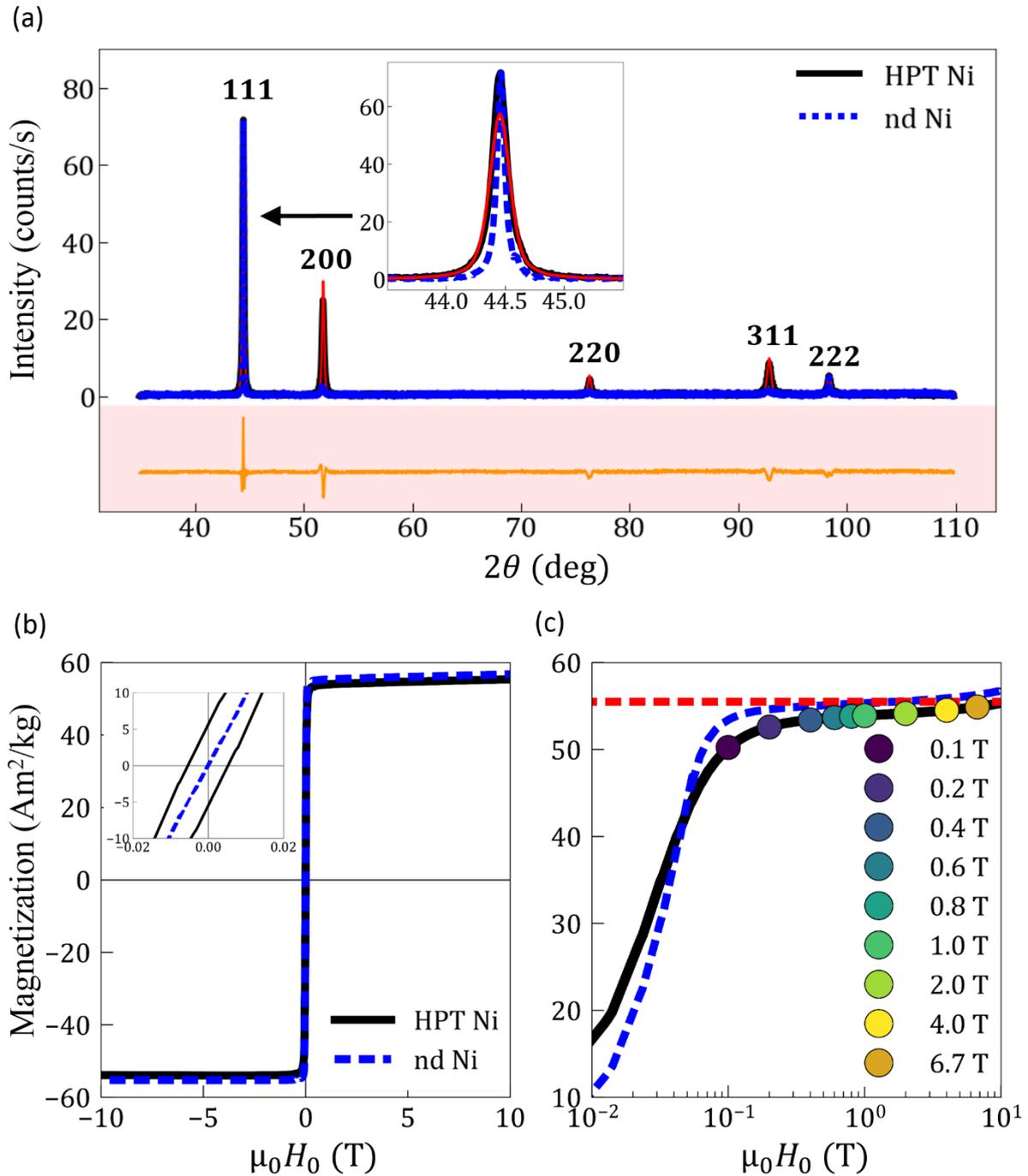

Figure 1: (a) X-ray diffractogram of HPT Ni (black solid line) and of nondeformed (nd) Ni (blue dashed line) (Cu K$_\alpha$ radiation). Red solid line: XRD data refinement using the Le Bail fit method implemented in the Fullprof software. The bottom orange solid line represents the difference between the calculated and observed intensities. (b) Room-temperature magnetization curves of HPT Ni and of nd Ni. Inset in (b): zoom of the magnetization curves between ± 20 mT. (c) Semi-logarithmic plot of the upper right branch of the magnetization curves plotted in Fig. 1(b). The large data points in (c) correspond to the field values (see inset) where the SANS measurements have been carried out. Horizontal dashed line: saturation magnetization $M_s$ = 55.5 Am$^2$/kg of bulk Ni [39].

# Figure 2

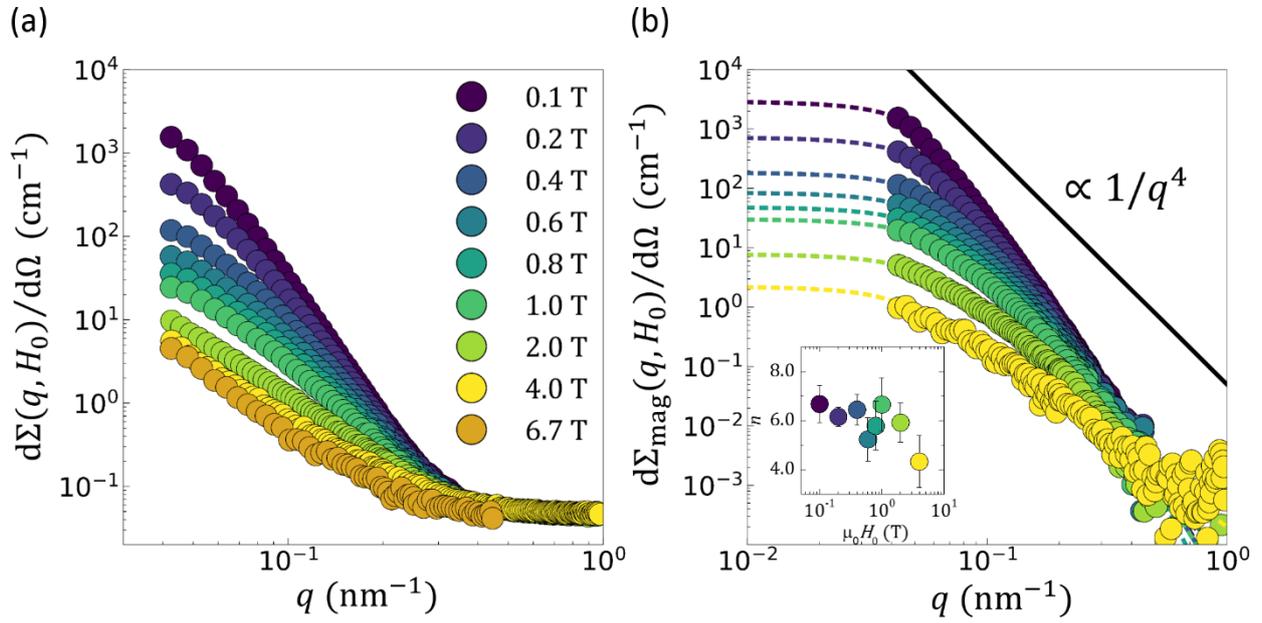

Figure 2: (a) Magnetic-field dependence of the (over 2π) azimuthally-averaged total (nuclear + magnetic) SANS cross section $d\Sigma(q, H_0)/d\Omega$ and (b) of the purely magnetic SANS cross section $d\Sigma_{mag}(q, H_0)/d\Omega$ (log-log scale). Dashed lines in (b): Extrapolation of $d\Sigma_{mag}(q, H_0)/d\Omega \propto 1/q^n$ from $q_{max}$ to infinity and of $d\Sigma_{mag}(q, H_0)/d\Omega \propto a + bq^2$ from $q_{min}$ to $q = 0$. Inset in (b): Field dependence of the asymptotic power-law exponent $n$ of the magnetic SANS cross section $d\Sigma_{mag}(q, H_0)/d\Omega$ on a semi-logarithmic scale.

# Figure 3

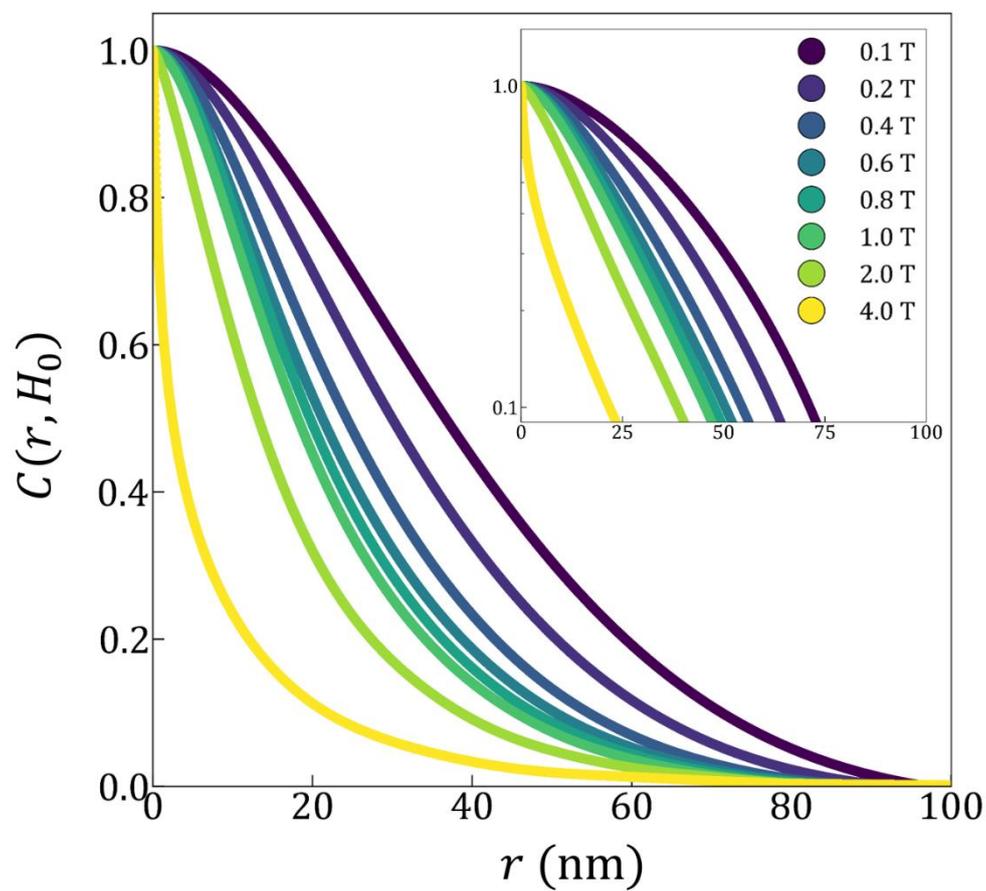

Figure 3: Magnetic-field dependence of the normalized magnetic correlation function $C(r, H_0)$. The correlation functions were numerically computed by a direct Fourier transformation [Eq. (3)] of $d\Sigma_{\text{mag}}(q, H_0)/d\Omega$ shown in Fig. 2(b). Inset: Plot of $C(r, H_0)$ on a semi-logarithmic scale, emphasizing the non-exponential decay of the correlations.

Figure 4

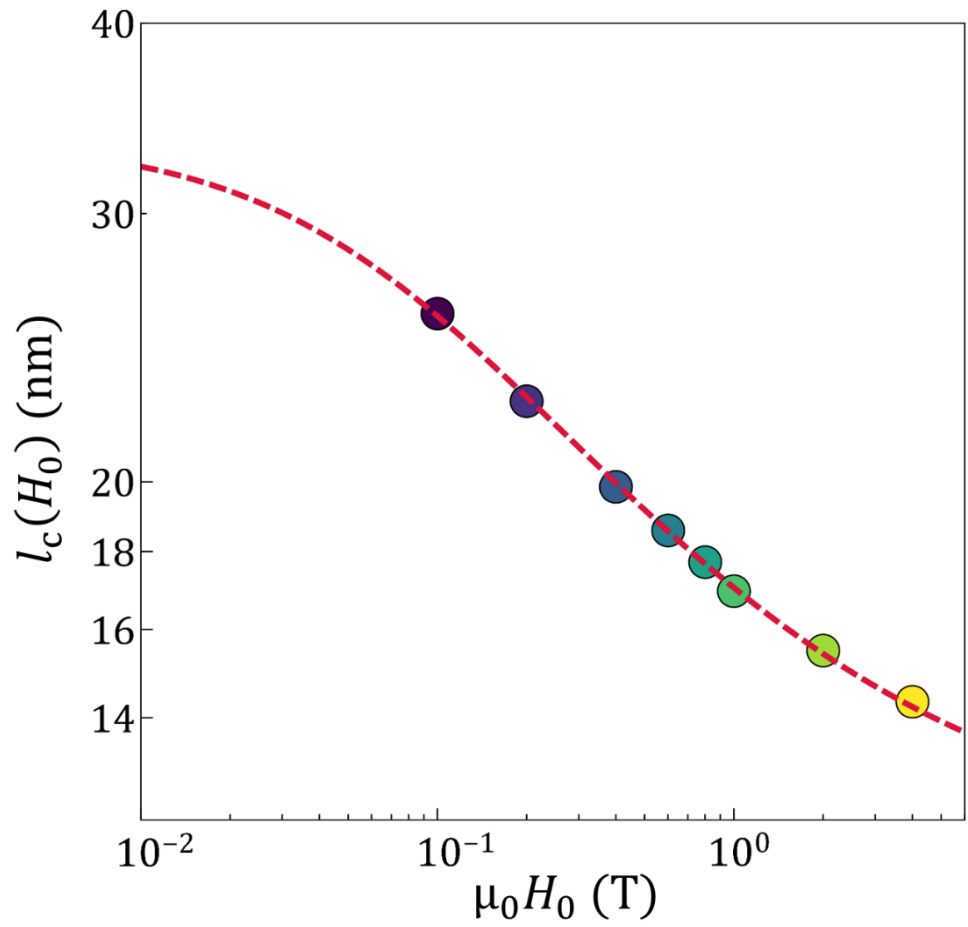

Figure 4: Field dependence of the magnetic correlation length $l_c(H_0)$, determined from the computed $C(r, H_0)$ data shown in Fig. 3 (log-log scale). Dashed line: Fit of the $l_c(H_0)$ data using Eq. (5).

# Figure 5

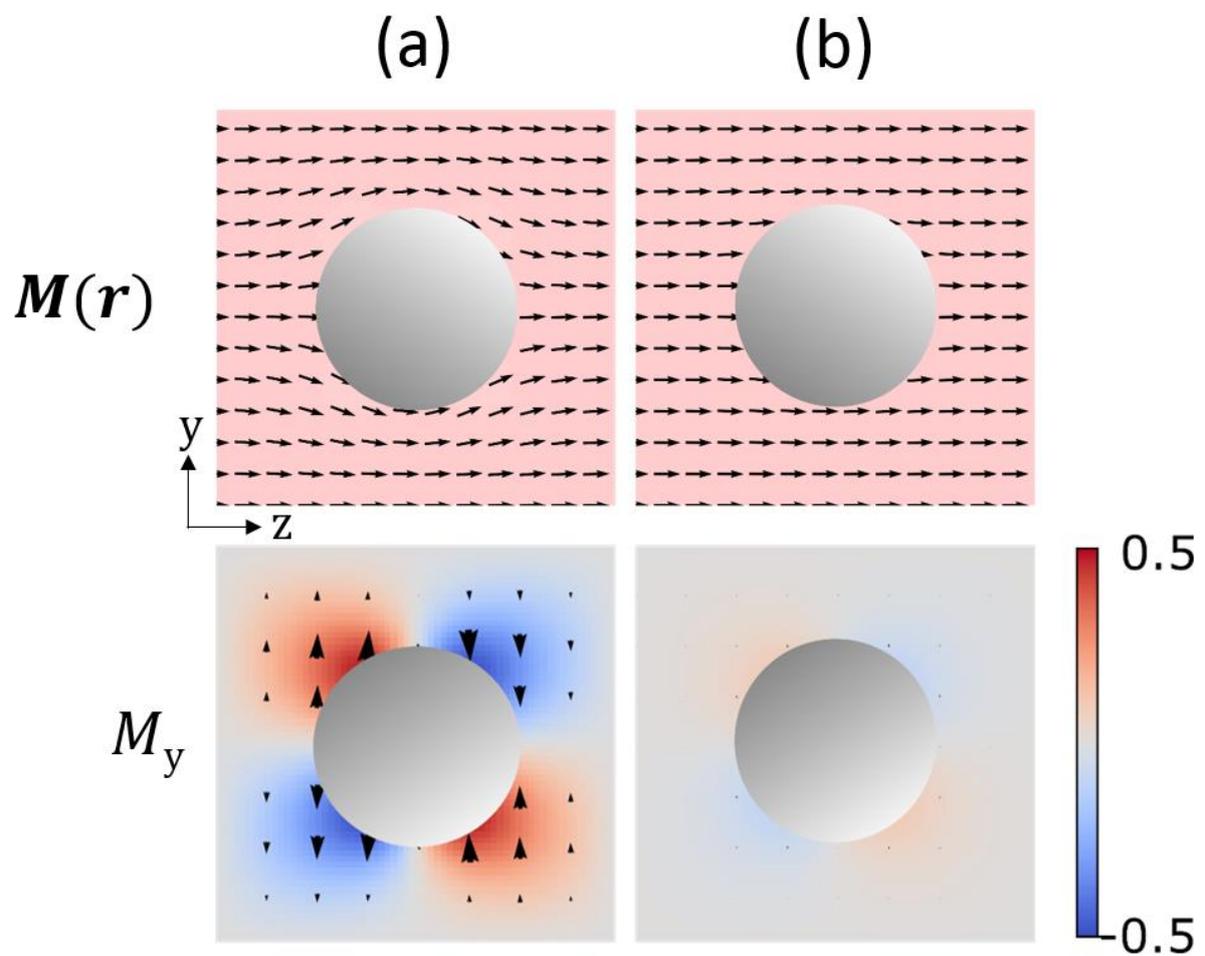

Figure 5: Illustration of the spin-misalignment correlations around defects. Shown are the results of micromagnetic simulations for a spherical defect (magnetic hole) embedded in a uniform Ni matrix. Applied-field values ($H_0$ is parallel to the $z$-direction): (a) 0.05 T and (b) 2 T. Top panel: projections of the three-dimensional magnetization distribution $M(r)$ into the $y$-$z$-plane. Bottom panel: perpendicular magnetization component $M_y$. The black arrows and color bars in (b) indicate the strength and orientation of $M_y$ (arbitrary units). Materials parameters of Ni were used (Ref. [40]).